# DCMD: Distance-based Classification Using Mixture Distributions on Microbiome Data

Konstantin Shestopaloff[1†], Mei Dong[1†], Fan Gao[1], Wei Xu[1,2*]

1. Dalla Lana School of Public Health, University of Toronto, 155 College Street, M5T 1P8, Toronto, CANADA

2. Princess Margaret Cancer Centre, University Health Network, 610 University Avenue, M5G 2M9, Toronto, CANADA

## Abstract

Current advances in next generation sequencing techniques have allowed researchers to conduct comprehensive research on microbiome and human diseases, with recent studies identifying associations between human microbiome and health outcomes for a number of chronic conditions. However, microbiome data structure, characterized by sparsity and skewness, presents challenges to building effective classifiers. To address this, we present an innovative approach for distance-based classification using mixture distributions (DCMD). The method aims to improve classification performance when using microbiome community data, where the predictors are composed of sparse and heterogeneous count data. This approach models the inherent uncertainty in sparse counts by estimating a mixture distribution for the sample data, and representing each observation as a distribution, conditional on observed counts and the estimated mixture, which are then used as inputs for distance-based classification. The method is implemented into a $k$-means and $k$-nearest neighbours framework and we identify two distance metrics that produce optimal results. The performance of the model is assessed using simulations and applied to a human microbiome study, with results compared against a number of existing machine learning and distance-based approaches. The proposed method is competitive when compared to the machine learning approaches and showed a clear improvement over commonly used distance-based classifiers. The range of applicability and robustness make the proposed method a viable alternative for classification using sparse microbiome count data.

---

*Correspondence: Wei.Xu@uhnresearch.ca

[†]Konstantin Shestopaloff and Mei Dong contributed equally to this work.

# 1   Introduction

The increasing accessibility of high-throughput technology has generated a wide array of different data types for analysis. One type of data that has recently gained in popularity is microbiome community data, composed of site specific counts for identified bacteria and notable for an abundance of sparse counts. A steadily growing number of studies have demonstrated associations between the human microbiome and health outcomes, such as inflammatory bowel disease [1], type 2 diabetes [2], cardiovascular disease [3], making it an important topic of research. However, the presence of sparsity and skewness, which characterizes this type of data, presents a number of challenges to statistical modelling since not all methods are designed to account for these features. This has motivated methodological developments to focus on expanding the existing set of approaches, particularly for classification tasks related to disease risk.

A variety of machine learning methods to predict disease outcome using microbiome data have been applied. A popular class of approaches have been distance-based methods, which differentiate and classify samples based on the distances derived from multivariate measures. Ubiquitous methods include $k$-means [4] and $k$-Nearest Neighbours ($k$-NN) [5] which have been adapted to such data with variable transformations and Euclidean distances, Manhattan distance, and a variety of other measures [6-8]. Other adaptations also include the distance-based nearest shrunken centroid (NSC) classifier, developed for use in microarray data [9]. It has also been utilized for microbiome data by taking the average of the relative abundance for each class from the training data as the class centroid [10, 11]. This classification is determined by the minimal standardized squared distance between the new sample and the class centroid.

A number of linear and additive machine learning classifiers are also commonly used for high-throughput data; LASSO, ridge regression (RR), gradient boosting (GB) and random forest (RF) are popular choices [7, 11-13]. Those methods rely on penalization (LASSO and RR) in logistic or multiclass logistic regression, typically with a log-transform of adjusted counts or relative abundances to address sparseness and skewness [14]. The GB and RF algorithms rely on sequentially constructed classifiers and automatically incorporate feature selection. However, none of these methods specifically address some of the issues arising from sparse count data. Other recent methodological developments for microbiome data include linear regression models with a phylogenetic tree-guided penalty term [15] and inverse regression to deal with the over-dispersion of zeros in count data [16]. Those methods have their limitations; for example, the tree-guided method can be overly influenced by tree information [15].

The existing methods all incorporate observed count data or relative abundance when computing distances or defining covariates, however, they do not explicitly account for the underlying uncertainty caused by sparsity in the observed data. In this paper, we aim to address these problems in a classification framework where predictors are sparse and heterogeneous count data. We present an innovative approach for distance-based classification using mixture distributions (DCMD) that specifically addresses the uncertainty in sparse and low count data and incorporate it into a distance-based classification framework. The method is tested using simulation studies and its effectiveness is illustrated by implementing into a human microbiome study [17]. The paper concludes with a discussion of the merits, drawbacks and the scope of applicability of the proposed methodology.

## 2  Method

In this section, we outline the DCMD method. The main steps of the model include mixture distribution specification and estimation for modelling observed data, calculation of conditional distributions for each sample, and calculating distances between samples and cluster centres to use in distance-based classification methods. The mixture model and conditional distribution estimation are described in Shestopaloff et al. [18]. It is proposed to model the underlying population rate structure of count data using a mixture distribution with Poisson-Gamma components, then conditioning on observed sample counts and resolution to obtain sample-specific distributions. An illustration of this procedure is shown in Fig. 1. In the next step, we use the sample-specification distributions for classification by calculate the distances between distributions instead of between counts or relative abundances. We incorporate DCMD into $k$-means and $k$-NN classification framework, which is shown in Fig. 2.

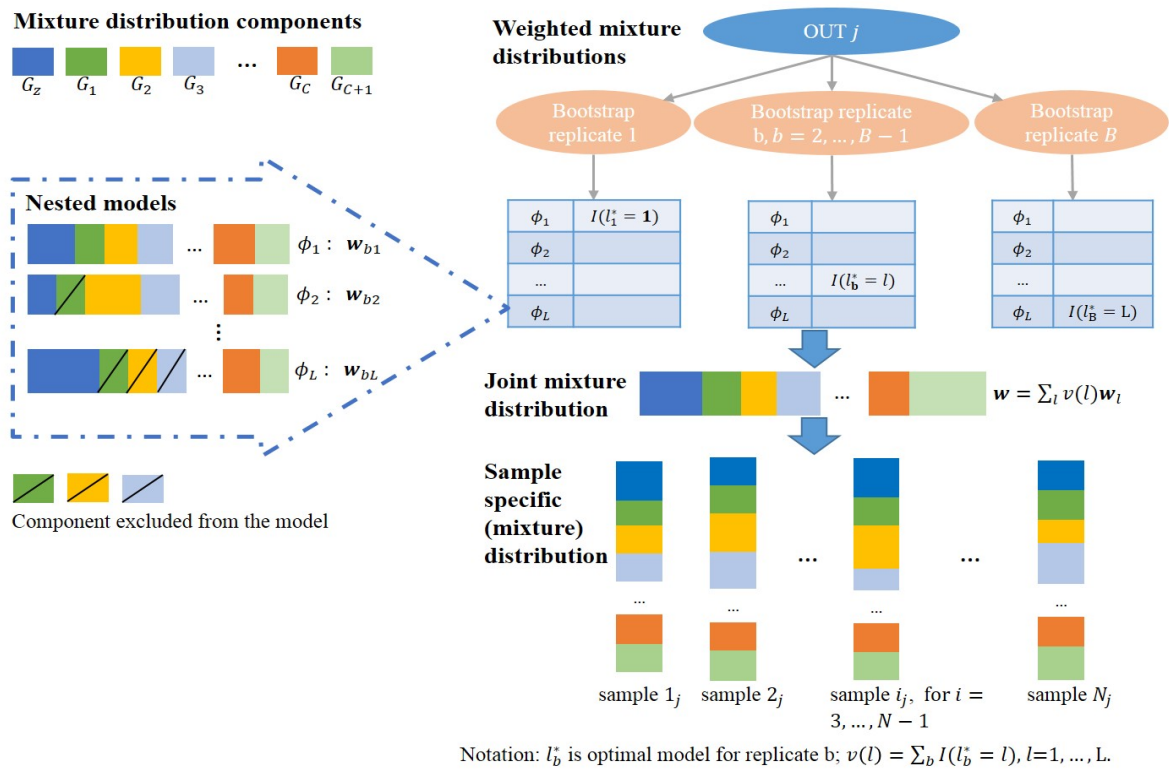

**Fig. 1 Workflow for obtaining a sample-specific mixture distribution for each sample i in OTU j.** The process is: 1) Specify a set of nested candidate mixture distributions using a specific set of components; 2) For each OTU j = 1, …, J, apply bootstrap to the set of nested models and calculate the weights of each candidate mixture model, then calculate the joint mixture distribution; 3) Estimate sample-specific distributions based on the joint mixture distribution.

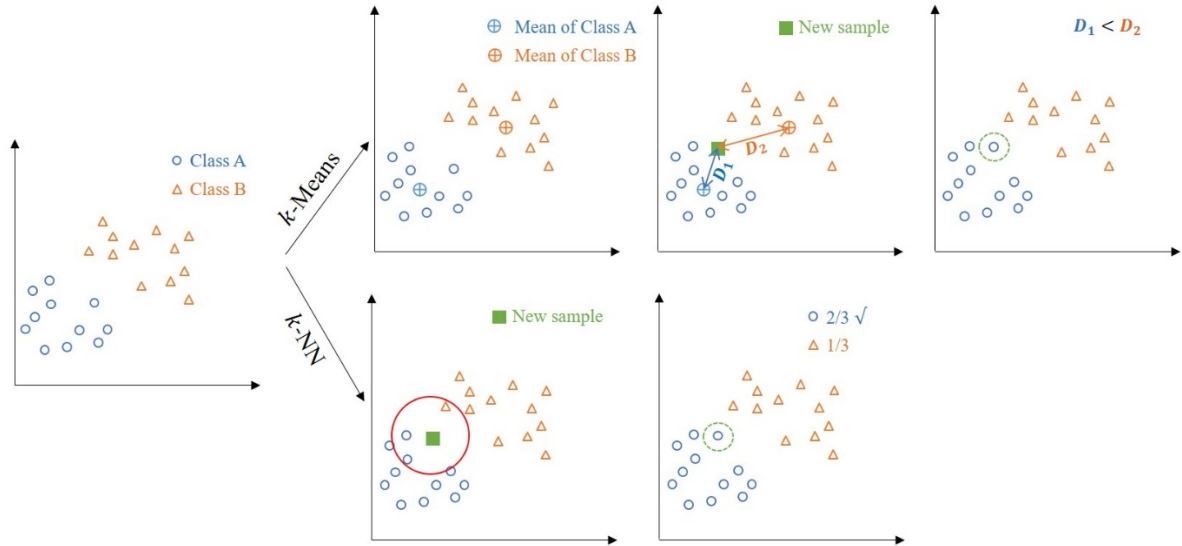

**Fig. 2 Illustration of $k$-means and $k$-NN framework using sample-specific distributions.** For $k$-means (top panel), the distance between the new sample to mean of class A is smaller than that to the mean of class B, hence the new sample is predicted as class A. For $k$-NN (bottom panel), using 3 nearest neighbours, the new sample is predicted as class A.

## 2.1 Model Specification and Estimation

Typical data in microbiome study consists of counts of operational taxonomic units (OTUs), as shown in Table 1. The notations used in our method formulation are as follows:

- $n_{ij}$, $i = 1, \ldots, I$ for $j = 1, \ldots, J$, is define the count of the $j$th OTU of the $i$th sample.

- $N_i$, the total number of aligned reads of sample $i$, $N_i = \sum_j n_{ij}$.

In order to model the underlying rate structure of the observed count data, we assume that the observed counts are Poisson distributed with rate

$$r_i = q_i N_i$$

for each OTU $j$ of sample $i$, where $q_i$ is individual-specific relative abundance and is sampled from some general population distribution $G_q$. Without loss of generality, we focus on a specific OTU and omit the $j$th subscript for subsequent notation. Then we have

$$r_i = q_i N_i = r_i^* t_i,$$

where $t_i = N_i/\bar{N}$, $r_i^* = q_i \bar{N}$ and $\bar{N} = \sum_i N_i / I$, with $r_i^*$ sampled from $G = G_q \bar{N}$, which is a rate normalized to the average sample reads, so the counts are treated on the same scale. Thus, the observed count for a specific site is

$$n_i | t_i \sim \text{Poisson}(r_i^* t_i),$$

where $r_i^* \sim G$. Since the distribution of OTU is zero-inflated, highly skewed, and heavy tailed, we propose a mixture distribution to approximate $G$. For positive rates on a given interval, we specify a set of Gamma components with integer $\alpha$ spaced uniformly on a log-scaled interval and $\beta = 1$, which corresponds to the rate posterior of a Poisson count. To reflect individuals that are never exposed to a specific OTU, we include a zero point mass, $n_i | t_i \sim 0$ where $P(n_i = 0) = 1$, to model the structural zeros. In addition, for sparse high rates, we define a high-count point mass, $n_i | t_i \sim C \cdot \mathbf{1}(n_i > C)$ where $P(n_i > C) = 1$, where $C$ is the truncation point and $\mathbf{1}(\cdot)$ is the indicator function. The final set of mixture components is $\mathbf{\Omega} = (G_z, G_1, G_2, \ldots, G_M, G_{C+})$, where $G_z$ is zero point mass, $G_m$, $m = 1, 2, \ldots, M$, is a set of Gammas with shape and rate parameters $(\alpha_m, \beta_m)$, and $G_{C+}$ is a high count point mass.

Table 1: A general OTU table for Microbiome data.

|  | **OTU 1** | ... | **OTU J** | **Total Reads** |
|---|---|---|---|---|
| Sample 1 | $n_{11}$ | ... | $n_{1j}$ | $N_1$ |
| ⋮ | ⋮ |  | ⋮ |  |
| Sample I | $n_{I1}$ | ... | $n_{IJ}$ | $N_I$ |

Define the weight of each component as

$$\mathbf{w} = (w_z, w_1, w_2, \ldots, w_M, w_{C+})',$$

where $w_z$ is the weight of zero point mass, $w_m$ is the weight for $m$th corresponding Gamma components, and $w_{C+}$ is the weight of high count point mass. Define

$$y_x = \sum_i \mathbf{1}(n_i = x),$$

the number of species observed $x$ times, then the weights are estimated by minimizing the sum of squares between observed and expected aggregate counts $y_x$ for $x = 0, 1, 2, \ldots, C, C+$. Note that given $\Gamma(\alpha_m, \beta_m)$, sample counts conditional on $t_i$ are distributed as a negative binomial distribution $NB[\alpha_m, \beta_m/(t_i + \beta_m)]$. Define

$$p_{xmi} = P_{NB}(X = x | t_i, \alpha_m, \beta_m)$$

as the probability of observing count $x$ from the $m$th mixture component conditional on the resolution $t_i$, then the expected aggregate count of $y_x$ is $\sum_{w_m \in w} w_m p_{xm} \cdot I$, where $p_{xm} = \sum_i p_{xmi}/I$. Thus, we have the objective function:

$$\arg\min_{\boldsymbol{w}} \sum_{x=0}^{C+} \left[ y_x - \left( \sum_{w_m \in w} w_m p_{xm} \right) \cdot I \right]^2, \tag{1}$$

$$s.t. \sum_m w_m = 1, w_m \geq 0, \forall m.$$

The estimate of $\boldsymbol{w}$ is obtained by optimizing the objective function (1).

## 2.2 Weighted Mixture Distribution

To address the uncertainly around specifying components for the low rates caused by sparsity, we define a set of nested models $\Phi_l, l = 1, \ldots, L$, with varying components for the low rate structure. We estimate the joint mixture model using a nonparametric bootstrap algorithm. As stated in Shestopaloff et. al. [18], we can obtain the weight $v(l)$ of each model, which is the proportion of times each model was selected as optimal relative to the observed data, then get the joint mixture distribution estimate. Let $\boldsymbol{w}_l$ be the estimated weights for each candidate model, $\Phi_l$, with zeros assigned to components not included in a specific model, then the weights of the final model are $\boldsymbol{w} = \sum_l v(l) \boldsymbol{w}_l$.

## 2.3 Sample Specific Distribution

By conditioning on the estimated weights, observed count $n_i$ and resolution $t_i$, we can obtain the probability that sample $i$ belongs to a specific component, as follows:

$$p_{im} = P(i \in G_m | n_i, t_i, \mathbf{w}) = w_{G_m} \frac{\Gamma(n_i + \alpha_m)}{\Gamma(n_i + 1)\Gamma(\alpha_m)} \left(\frac{\beta_m}{t_i + \beta_m}\right)^{\alpha_m} \left(1 - \frac{\beta_m}{t_i + \beta_m}\right)^{n_i}. \tag{2}$$

The probability of being assigned to the two point masses is $P(i \in G_0) = \mathbf{1}(n_i = 0)$ and $P(i \in G_{C+}) = \mathbf{1}(n_i > C)$. The sample specific mixture weights are

$$\mathbf{w}_i = (w_{iz}, w_{i1}, \dots, w_{iC}, w_{iC+})',$$

where

$$w_{im} = P(i \in G_m) / \sum_m P(i \in G_m) = p_{im} / \sum_m p_{im}.$$

Note that since component assignment probabilities have been adjusted for individual resolution, the Poisson scaling factor is unity, so the Poisson-Gamma mixture probabilities are $NB(\alpha, \beta/(1 + \beta))$. Also note that we have differentiated the zeros in our mixture distribution, define structural zeros as $x = z$ and observed zeros as $x = 0$. Therefore, given the underlying rate distribution with the joint mixture model, we can calculate the probability of observing count $x = z, 0, 1, \dots, C, C+$ from each mixture component, $G_m$, as

$$P_{G_m}(x) = P(X = x | G_m) = P_{NB}(X = x | \alpha_m, \beta_m).$$

For point mass, we have $P(X = x | G_z) = \mathbf{1}(n_i = 0)$ and $P(X = x | G_{C+}) = \mathbf{1}(n_i > C)$, respectively. Define a vector of probabilities

$$\mathbf{P}(x) = [P(X = x | G_z), P(X = x | G_1), \dots, P(X = x | G_m), P(X = x | G_{C+})]$$

$$= [P_{G_z}(x), P_{G_1}(x), \dots, P_{G_M}(x), P_{G_{C+}}(x)],$$

for $= z, 0, 1, \dots, C, C+$. The probability of observing $G_z$ is $P_{G_z}(z) = 1$ and 0 otherwise. The probability of observing $G_{C+}$ is $P_{C+}(C+) = 1$ and 0 otherwise.

Then we can define the PDF for sample $i$ as

$$\mathbf{P}_i = [P_i(z), P_i(0), \dots, P_i(C), P_i(C+)] = \mathbf{w}_i' \mathbf{P},$$

where

$$P = [P(z), P(0), \ldots, P(C), P(C+)],$$

$$P_i(z) = w_{G_z},$$

$$P_i(C+) = 1 - [\sum_{x=0}^{C} P_i(x) + w_{G_z}].$$

The matrix $P$ is a convenient construct as it can be pre-calculated for distance calculations, since it will be the same for all samples.

## 2.4 Classification

In this section, we introduce our classification algorithm, DCMD, and outline how it can be used in a $k$-means and $k$-NN classification framework by incorporating distance-based measures, specifically, discrete $L^2$-PDF norm and continuous $L^2$-CDF norm.

### 2.4.1 Distances

Given a mixture distribution CDF $F_i$, PDF $f_i$ and an estimated set of weights $w_i$ for sample $i$, the metrics used are:

**Discrete $L^2$-PDF norm**

$$\delta_{PDF}(f_i, f_j) = \sum_x [f_i(x) - f_j(x)]^2$$
$$= \sum_x [(w_i - w_j) P(x)]^2, \quad (3)$$

where $x = z, 0, 1, \ldots, C, C+$. Note that we include the structural zero component, $z$, separately and that the distances only depend on the weights. For multiple predictors, $j$, the total distance between samples $i_1$ and $i_2$ is the sum across all predictors, $D(i_1, i_2) = \sum_j \delta_j(f_{i_1}, f_{i_2})$.

**Continuous $L^2$-CDF norm**

$$\delta_{CDF}(F_i, F_j) = \int_0^C [F_i(x) - F_j(x)]^2 \, dx$$

$$= (w_i - w_j)G_{m_1 m_2}(w_i - w_j)', \tag{4}$$

where $G_{m_1 m_2}$ is a matrix where $(m_1, m_2)$ entry is $\int G_{m_1}(x)G_{m_2}(x)\,dx$ for each of the continuous mixture component. Details of the derivation can be found in Shestopaloff [19].

### 2.4.2 Distance Based Classification

We implement our distance calculated above into $k$-means and $k$-NN algorithm. The workflow is shown in Fig. 2.

**K-means:** To adapt to the $k$-means algorithm, we need to estimate a distribution for the mean of each class by minimizing the distributional distances between it and the class samples, conditional on a specified distance. Since the $L^2$ norms only depend on the weights, as shown in Equation (3) and (4), we only need to determine the mean of the weights for each class. The algorithm is implemented as follows:

Step 1: Determine the mean of weight for the $j$th predictors in class $k$,

$$w_{\mu_{k,j}} = \sum_{i \in k,j} w_{k,j} / |N_{k,j}|,$$

where $|N_{k,j}|$ is the number of samples in class $k$ of predictor $j$, $k = 1, \dots, K$ and predictor $j = 1, \dots, J$;

Step 2: Compute the overall distance to the mean for sample $i$ across all predictors,

$$D(i, \mu_k) = \sum_j \delta(P_{i,j}, P_{\mu_k,j});$$

Step 3: The label of sample $i$ is predicted as

$$\hat{y}_i = \underset{k}{\mathrm{argmin}}\, D(i, \mu_k).$$

**K-NN:** For $k$-NN, after computing the pairwise distances between samples and summing across predictors, these can be directly used to identify the nearest neighbours for classification. The algorithm of $k$-NN is described as follows:

Step 1: Compute the pairwise distance of sample $i_1$ and $i_2$, $i_1, i_2 = 1, \dots, I$, $i_1 \neq i_2$,

$$D(i_1, i_2) = \sum_j \delta(P_{i_1,j}, P_{i_2,j});$$

Step 2: For sample $i$, pick the $k$ samples with smallest distance to sample $i$, the optimal $k$ can be determined using cross-validation in the training set or existing heuristics.

Step 3: Tally the labels of the $k$ nearest neighbours, then sample $i$ is predicted as the mode of the $k$ labels.

## 2.5 Predictive Metrics

Let $\hat{y}_i$ is the predictive class for sample $i$ obtained from $k$-means or $k$-NN. The accuracy is defined as the proportion of correctly predicted cases: $ER = \frac{1}{I}\sum_{i=1}^{I} I(\hat{y}_i = y_i)$. For binary outcomes we also include precision, recall, and F1 score as metrics to measure the predictive performance. We count the number of true positive ($TP$), false positive ($FP$), and false negative ($FN$), then precision, recall, and F1 score are defined as follows [20]:

$$\text{precision} = TP/(TP + FP),$$

$$\text{recall} = TP/(TP + FN),$$

$$\text{F1 score} = 2 * \text{precision} * \text{recall}/ (\text{precision} + \text{recall}).$$

# 3 Simulation

## 3.1 Data generation

To evaluate the performance of the DCMD method, we simulate data to mimic microbiome community count data and assess classification accuracy. We specify 3 classes and measure the performance for six simulation scenarios, averaging over 100 replications. Different combinations of signals and zero proportions (ZP) are generated to examine model performance, along with a null case. The sample size of each class is $N_{1,j} = N_{2,j} = N_{3,j} = 400$ for $j = 1, \ldots, J$, and $J = 25$. Each OTU count is sampled from a Poisson-Gamma mixture with the rate distribution being a mixture of Gammas and a structural zero component. A null case scenario is generated by permuting class labels.

The number of counts in each mixture component is set by binning samples from a $Beta(\alpha_b, \beta_b)$ at uniform intervals, with the $\alpha_b$ varied to result in different class means and levels of sparsity, and $\beta_b \sim Unif(2, 6.5)$. The specified number of rates, $r^*$, are sampled from each of the components. The number of components $M$ is sampled from $Unif(5, 15)$ and the data range for scaling is sampled from a $Unif(100, 300)$. The resolution $t_i$ for each observation is sampled from $Unif(2/3, 4/5)$ and standardized across the sample, observed counts are $n_i \sim Poisson(r_i^* t_i)$.

The parameters for each class and the corresponding ZP, standard deviation and mean count for each scenario are shown in Table 2. Scenarios 1, 2, and 5 represent low ZP; scenarios 3, 4, and 6 represent high ZP, with scenario 6 representing the null case.

### 3.2 Model Fitting and Comparison Methods

The proposed method is compared with $k$-means and $k$-NN using Euclidean and Manhattan distances of relative abundances as well as the distance-based NSC. Additionally, we compare our method with classifiers using LASSO, RF, GB and RF. Models are trained using a 60/40 training/test set split [21]. The training set remains the same for all classifiers within each replication. The model specifications for each classifier are reported in Supplementary S1 and S2 (Additional file). We use predictive accuracy to compare performance.

The simulations and methods implementation was conducted based on in R v.3.5.0; NSC is implemented *pamr* [22], RF in *randomforest* [23], GB in *gbm* [24], and LASSO and RR in *glmnet* [25].

### 3.3 Simulation Results

The classification accuracy of each model and scenario is presented in Fig. 3. We also show the summary of accuracy of each model in Table 3. We calculate the accuracy of each model on the test set across 100 replicates.

From the results, it is obvious that within $k$-means and $k$-NN classifiers DCMD using $L^2$-PDF and $L^2$-CDF norms outperforms regular $k$-means and $k$-NN classifiers using Euclidean or

Table 2: The parameter setting of each scenario and the corresponding ZP and mean count for each class over 100 replicates.

| Scenario | Signal | Sparsity | Class | $\alpha_b$ | Mean ZP (SD) | Mean |
|---|---|---|---|---|---|---|
| 1 | Weak | Low | 1 | (1.6, 2.0) | 0.30 (0.13) | 8.51 |
| | | | 2 | (2.0, 2.4) | 0.23 (0.11) | 10.86 |
| | | | 3 | (2.4, 2.8) | 0.17 (0.10) | 13.41 |
| 2 | Strong | Low | 1 | (1.2, 1.8) | 0.37 (0.14) | 6.99 |
| | | | 2 | (2.0, 2.4) | 0.23 (0.11) | 11.04 |
| | | | 3 | (2.6, 3.0) | 0.15 (0.09) | 14.83 |
| 3 | Weak | High | 1 | (0.4, 0.8) | 0.70 (0.12) | 3.13 |
| | | | 2 | (0.6, 1.0) | 0.61 (0.13) | 3.17 |
| | | | 3 | (0.8, 1.2) | 0.53 (0.14) | 3.24 |
| 4 | Strong | High | 1 | (0.2, 0.3) | 0.87 (0.06) | 0.85 |
| | | | 2 | (0.6, 0.7) | 0.67 (0.11) | 2.52 |
| | | | 3 | (1.0, 1.1) | 0.50 (0.14) | 4.49 |
| 5 | Strong | Extreme high | 1 | (0.1, 0.2) | 0.92 (0.04) | 1.16 |
| | | | 2 | (0.25, 0.35) | 0.84 (0.07) | 1.17 |
| | | | 3 | (0.5, 0.6) | 0.72 (0.10) | 1.18 |
| 6 | Null | Low | 1 | (1.2, 1.8) | 0.37 (0.14) | 10.98 |
| | | | 2 | (2.0, 2.4) | 0.23 (0.11) | 10.91 |
| | | | 3 | (2.6, 3.0) | 0.15 (0.09) | 10.97 |

Manhattan distances. The mean accuracy of $k$-means or $k$-NN based on Euclidean or Manhattan distance is around 0.41 when the signal is weak (scenarios 1 and 3), which is a nominal improvement compared to a baseline accuracy of 0.33. In contrast, $k$-means with $L^2$-norms achieves accuracies of around 0.60 and $k$-NN with $L^2$-norms around 0.50. In scenarios with stronger signals (scenarios 2, 4, and 5) the differences in accuracy between DCMD and regular $k$-means and $k$-NN are even more pronounced. Additionally, DCMD is more stable than regular $k$-means and $k$-NN, as the standard deviations (SDs) of the former are notably smaller, as shown in Table 3.

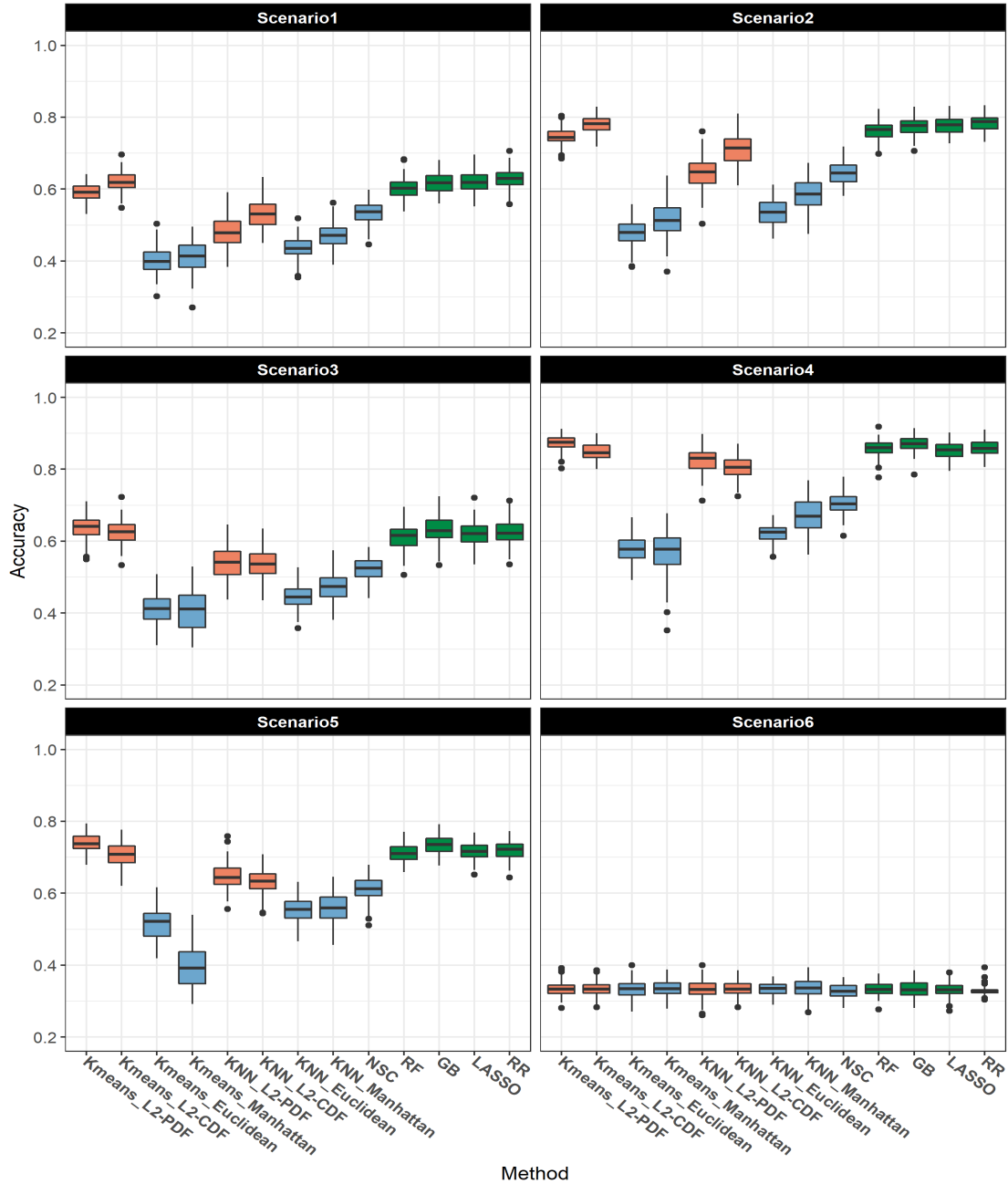

**Fig. 3 Boxplot of the accuracy over 100 replicates for each method in each scenario.** The proposed DCMD method is shown in red for $k$-means and $k$-NN methods with L2-PDF and L2-CDF distances. Comparison distance-based methods are shown in blue and include $k$-means and $k$-NN with Euclidean and Manhattan distances and NSC. Machine learning methods are shown in green and include random forest (RF), gradient boosting (GB), LASSO, and ridge regression (RR).

Compared to DCMD within $k$-NN framework, we can also see that the accuracies of DCMD within $k$-means framework are higher. It is also interesting to note that DCMD using $L^2$-PDF norm performs slightly better than continuous L^2-CDF norm when ZP is low, but is the opposite when ZP is higher. Nearest shrunken centroid performs close to DCMD within k-NN framework, but falls short of DCMD performance within k-means and other machine learning methods.

In addition, we observe that our proposed classifier is comparable to the standard machine learning methods (RF, GB, LASSO, RR) across simulation scenarios. The optimal distance-based classifier and optimal machine learning method are highlighted in scenario 1 to 5. When ZP is low in scenario 1 and 2, the performances of DCMD within $k$-means are close to the machine learning methods. When ZP is high in scenario 3 and 4, $k$-means with $L^2$-PDF norm performs slightly better than the top machine learning method. When the data is extremely sparse, as in scenario 5, DCMD outperforms the four machine learning methods and other distance-based classifiers. We can also notice that the performance of DCMD is very stable across all scenarios when applied in a $k$-means framework. The classifier also performs as expected in the null case, predictive accuracy of each model is close to 0.33.

## 4  Application

### 4.1  Data Description

We apply our method to a microbiome study on colorectal cancer reported by [17]. A total of 190 samples (95 pairs) were collected from 95 patients in Vall d'Hebron University Hospital in Barcelona and Genomics Collaborative. The study aimed to identify associations between tumor microbiome and colorectal carcinoma. Both the colorectal adenocarcinoma tissue and adjacent non-affected tissues were obtained from subjects. The OTU count table generated by 16S amplification was obtained from Microbiome Learning Repo [12]. Prior to model training, eighteen samples with total reads less than 200 were dropped from the dataset and we excluded OTUs with mean relative abundance of less than 0.001, resulting in 149 OTUs and 172 samples (86 pairs) used to differentiate tumour and normal tissue.

Table 3: The summary of accuracy for each model in six scenarios over 100 replicates.

| Model | Scenario 1 (SD) | Scenario 2 (SD) | Scenario 3 (SD) | Scenario 4 (SD) | Scenario 5 (SD) | Scenario 6 (SD) |
|---|---|---|---|---|---|---|
| $k$-means-$L^2$-PDF | 0.59 (0.026) | 0.75 (0.024) | **0.64 (0.034)** | **0.87 (0.020)** | **0.74 (0.026)** | 0.33 (0.022) |
| $k$-means-$L^2$-CDF | **0.62 (0.027)** | **0.78 (0.023)** | 0.62 (0.035) | 0.85 (0.023) | 0.71 (0.031) | 0.33 (0.021) |
| $k$-means-Euclidean | 0.40 (0.041) | 0.48 (0.038) | 0.41 (0.040) | 0.58 (0.038) | 0.51 (0.043) | 0.33 (0.023) |
| $k$-means-Manhattan | 0.41 (0.041) | 0.51 (0.051) | 0.41 (0.055) | 0.57 (0.059) | 0.40 (0.057) | 0.33 (0.022) |
| $k$-NN-$L^2$-PDF | 0.48 (0.042) | 0.64 (0.047) | 0.54 (0.048) | 0.82 (0.034) | 0.65 (0.038) | 0.33 (0.026) |
| $k$-NN-$L^2$-CDF | 0.53 (0.043) | 0.71 (0.044) | 0.54 (0.042) | 0.80 (0.030) | 0.63 (0.035) | 0.33 (0.022) |
| $k$-NN-Euclidean | 0.44 (0.031) | 0.53 (0.036) | 0.45 (0.034) | 0.62 (0.027) | 0.55 (0.033) | 0.33 (0.019) |
| $k$-NN-Manhattan | 0.47 (0.035) | 0.59 (0.041) | 0.47 (0.038) | 0.67 (0.046) | 0.56 (0.038) | 0.34 (0.024) |
| NSC | 0.53 (0.031) | 0.65 (0.029) | 0.52 (0.033) | 0.71 (0.031) | 0.61 (0.036) | 0.33 (0.019) |
| RF | 0.60 (0.029) | 0.76 (0.027) | 0.61 (0.037) | 0.86 (0.023) | 0.71 (0.026) | 0.33 (0.018) |
| GB | 0.62 (0.029) | 0.77 (0.025) | 0.63 (0.037) | **0.87 (0.021)** | **0.73 (0.026)** | 0.33 (0.023) |
| LASSO | 0.62 (0.031) | 0.78 (0.023) | 0.62 (0.034) | 0.85 (0.023) | 0.72 (0.026) | 0.33 (0.021) |
| RR | **0.63 (0.028)** | **0.78 (0.022)** | **0.63 (0.036)** | 0.86 (0.021) | 0.72 (0.026) | 0.33 (0.011) |

### 4.2 Model Fitting and Evaluation

In order to reduce the predictor space for distance-based classifiers, we screen the OTUs using a non-parametric Mann-Whitney U Test to assess the association signals of each OTU on the training set. To adjust for multiple comparisons, we use q-values obtained by the Benjamini-Hochberger (BH) method [26] and retain OTUs with q-values less than 0.05 in each training set. The mean number of OTUs selected from each training set is 41.8 (range: 13-57). For machine learning approaches, we include all 149 OTUs. We compare our proposed $L^2$-based $k$-means and $k$-NN classifier with five other distance-based classifiers ($k$-means-Euclidean, $k$-means-

Manhattan, $k$-NN-Euclidean, $k$-NN-Manhattan, NSC) and four machine learning methods (RF, GB, LASSO, RR). We assess the model performance using 10-fold CV. For each iteration, one fold of the data is treated as the test set and the rest are used for training. The specification of DCMD and other classifiers is the same as for the simulations, given in Supplementary S1 (Additional file). In addition to accuracy, we calculate precision, recall, and F1 score as metrics to compare classifiers. The calculations of those metrics are done using functions in R Package *caret* [27].

## 4.3 Applied Results

The predictive performance of each classifier on the colorectal cancer study is shown in Table 4. The accuracy of $k$-means with $L^2$-PDF norm is 0.67, which is one the highest among all classifiers. The precision, recall, and F1 score are also favorable compared to other methods. For $k$-means with $L^2$-CDF norm, performance is comparable to some of the distance-based methods for this dataset, such as $k$-means with Euclidean distance and $k$-NN with Manhattan distance. Although NSC achieves accuracy with 0.67 and precision with 0.72, the recall and F1 score of NSC is much lower than $k$-means. The DCMD approach in a $k$-NN framework is inferior to $k$-means, as in the simulations. However, within the $k$-NN framework, the $L^2$ norms outperform Euclidean and Manhattan distances.

Compared to standard machine learning methods, DCMD is superior to RF, GB, LASSO, and RR for this dataset. The accuracy of RR is 0.64, following by GB with accuracy of 0.63, which are both smaller than DCMD with $L^2$-PDF norms. The predictive accuracies of RF and LASSO were not competitive in this particular dataset. We also check the performance of the machine learning methods using the OTUs selected for the distance-based classifiers (Supplementary Table S2) to ensure differences are not a result of variable selection.

Table 4: The predictive performance of the 13 classifiers on the colorectal cancer data.

| Method | Accuracy | Precision | Recall | F1 score |
|---|---|---|---|---|
| $k$-means-L$^2$-PDF | **0.67** | 0.66 | **0.69** | **0.67** |
| $k$-means-L$^2$-CDF | 0.63 | 0.62 | **0.67** | **0.65** |
| $k$-means-Euclidean | 0.62 | 0.62 | 0.60 | 0.61 |
| $k$-means-Manhattan | 0.65 | 0.66 | 0.60 | 0.63 |
| $k$-NN-L$^2$-PDF | 0.65 | **0.77** | 0.43 | 0.55 |
| $k$-NN-L$^2$-CDF | 0.63 | 0.65 | 0.57 | 0.61 |
| $k$-NN-Euclidean | 0.63 | 0.66 | 0.55 | 0.60 |
| $k$-NN-Manhattan | 0.61 | 0.69 | 0.40 | 0.50 |
| NSC | **0.67** | **0.72** | 0.56 | 0.63 |
| RF | 0.60 | 0.61 | 0.59 | 0.60 |
| GB | 0.63 | 0.63 | 0.64 | 0.64 |
| LASSO | 0.59 | 0.59 | 0.59 | 0.59 |
| RR | 0.64 | 0.63 | **0.69** | **0.66** |

## 5 Discussion

The results of our simulation study and applied example indicate that the proposed DCMD method performs well over a range of scenarios, achieving good classification performance when using sparse data as predictors. The predictive accuracy is improved compared to other distances within distance-based classifiers and it is either competitive or slightly advantageous relative to a number of standard machine learning methods under a wide range of scenarios.

The improved performance of DCMD on sparse data results from the use of mixture distributions to represent observed count data, as this allows for a separate zero inflation component and to better model the underlying uncertainty in the observed sample counts. This improved performance is observed in both the simulation study and real microbiome data, since this model formulation was better able to capture the underlying rate distributions in the data. This was particularly significant in comparison to other distances within the $k$-means and $k$-NN classifiers, where accuracy was consistently higher, in some cases up to +0.30, compared to standard distances using observed data.

The performance differences between the $L^2$-PDF and $L^2$-CDF norms can be attributed to data structure, as less sparse scenarios were better modelled by a rate structure, resulting in a slight advantage for the continuous $L^2$-CDF metric, while in the higher zero proportion and low-count scenarios, specific differentiation of zeros into structural and non-structural, as well as modelling expected counts directly, better captured the general structure of predictors used for differentiation.

In comparison to standard machine learning approaches, the model performances are generally on par, with slight improvements of DCMD in some cases. Among the four machine learning methods, we note that GB and RR tended to perform better and were more consistent than RF and LASSO in both simulations and real data, which may give us guidance that GB and RR are alternatives in microbiome classification problems.

As the DCMD method derives its major improvement from a focus on modelling lower count data, it is necessary to accurately specify an underlying set of mixture components for the low rates. At the same time the mixture has to model low and high count data on the same scale, where the latter is often sparse due to a lack of observations and it is not feasible to apply a transformation to make the data denser, due to abundant zeros and the discrete nature of low-counts. However, the suggested heuristic of specifying higher count distributions centered on a log-linear scale has worked well in our simulations, mimicking the log-transformation commonly applied to such data.

The proposed DCMD method is formulated in a distance-based framework so it does not include specific mechanisms for variable selection. While different predictors can alternatively be included in the distance sum, the process is not automated. In addition, the model specification is focused on count data using microbiome site counts, and continuous covariates need to be modelled separately using continuous distributions, while categorical covariates can only be included as dummy variables. These variables will also all be treated on the same scale in the distance metric, unless specified otherwise.

Despite some of these drawbacks, the representation of observations as distributions and the information inherent in the distance metric is able to compensate some of the disadvantages inherent in distance-based methods and achieve competitive performance in relation to more sophisticated classifiers as well as specially designed approaches like NSC. Overall, we have demonstrated the advantage of the DCMD approach for classification when data is expected to be sparse, particularly within a distance-based framework.

# 6   Conclusion

In this paper, we present a distance-based classification method for microbiome count data. The DCMD approach models the observed data using mixture distributions and calculates $L^2$-norms for use in distance-based classification algorithms. The method is specifically designed to accurately model low-count structures, addressing the inherent sparsity by representing each observed count as a distribution, and is demonstrated to have improved performance using simulation studies and a real microbiome study. The performance of the proposed DCMD is competitive to a number of machine learning methods and significantly outperforms other common metrics in distance-based classification models. The consistent and improved performances across a variety of different data structures make this approach a viable alternative for modelling and classification of microbiome count data.

**Abbreviations**

DCMD: distance-based classification using mixture distributions; *k*-NN: *k* Nearest Neighbours; NSC: nearest shrunken centroid; RR: ridge regression; GB: gradient boosting; RF: random forest; OTU: operational taxonomic unit; ZP: zero proportion; SD: standard deviation

# Supplementary Material

# DCMD: Distance-based Classification Using Mixture Distributions

Konstantin Shestopaloff[1†], Mei Dong[1†], Fan Gao[1], Wei Xu[1,2*]

1. Dalla Lana School of Public Health, University of Toronto, 155 College Street, M5T 1P8, Toronto, CANADA

2. Princess Margaret Cancer Centre, University Health Network, 610 University Avenue, M5G 2M9, Toronto, CANADA

**Supplementary information**

**S1: Specification of DCMD of simulated dataset**

The pre-specified minimum $\alpha_m$ cutoff is set to be 8 for the model initialization, and the maximum alpha quantile is 0.85. Five nested mixture models are pre-defined to fit the simulated data, in particular, $\Gamma(1, 2), \Gamma(1, 1), \Gamma(2, 1), \Gamma(3, 1), \Gamma(4, 1)$, by including all for the first model, all except $\Gamma(1,2)$ for the second, etc. In addition, the following are included for all models: $\Gamma(5, 1), \Gamma(6, 1), \Gamma(7, 1), \Gamma(8, 1)$. For each OTU, Gamma components with $\alpha_m$ specified on a linear-log scale from 8 to the 85% quantile of the OTU, a structural zero-point mass, and a high count point mass at the 85% quantile of the OTU are also included for all models. The models of one OTU in a dataset from scenario 1 are summarized in **Error! Reference source not found.**1. We estimate the joint model through 300 bootstrap replicates, with the $L^2$-PDF norm or continuous $L^2$-CDF norm used as the distance measure. The optimal estimates of weights are obtained by minimizing the least squares objective function using the Broyden-Fletcher-Goldfarb-Shanno algorithm [1] with the augmented Lagrangian method [2] for the constraints.

The estimation was done using R Package *NLopt* [3]. Once the estimates of weights for the joint model are obtained, we can estimate the class mean of each OTU in the training set.

**S2: Specification of Other Classifiers**

For $k$-means based on Euclidean distance and Manhattan distance, the distance between the new sample and class mean is calculated using normalized OTU relative abundances. The class mean is the average of normalized OTU for each class of the training set. For $k$-NN based on Euclidean distance and Manhattan distance, 10-fold cross-validation (CV) is implemented in the training data to find the optimal $k$. For NSC, OTUs are also normalized. For LASSO and Ridge Regression, we fit models on $log(x + 1)$ transformed OTU counts. We select the tuning parameter $\lambda$ with the minimal error rate from the training set using CV. For gradient boosting, the number of trees is chosen from 500, 600, 800, 900, and 1000. Other parameters are fixed. For random forest, we choose from 100, 500, or 1000 trees. The number of variables selected at a node split is chosen from integers 3 to 8. The best combination of the number of trees and the number of variables is returned through 10-fold CV using R Package *caret* [4].

**Supplementary Tables**

Supplementary Table S1: The nested models of mixture distribution components used in fitting one of the simulated data.

| Model/Included | Model 1 | Model 2 | Model 3 | Model 4 | Model 5 |
|---|---|---|---|---|---|
| $P(X = 0) = 1$ | X | X | X | X | X |
| $\Gamma(1, 2)$ | X | | | | |
| $\Gamma(1, 1)$ | X | X | | | |
| $\Gamma(2, 1)$ | X | X | X | | |
| $\Gamma(3, 1)$ | X | X | X | X | |
| $\Gamma(4, 1)$ | X | X | X | X | X |
| $\Gamma(5, 1)$ | X | X | X | X | X |
| $\Gamma(6, 1)$ | X | X | X | X | X |
| $\Gamma(7, 1)$ | X | X | X | X | X |
| $\Gamma(8, 1)$ | X | X | X | X | X |
| $\Gamma(11, 1)$ | X | X | X | X | X |
| $\Gamma(16, 1)$ | X | X | X | X | X |
| $P(X > 16) = 1$ | X | X | X | X | X |

Supplementary Table S2: The predictive performance of machine learning methods using the OTUs selected from Wilcoxon-rank sum test.

| Method | Accuracy | Precision | Recall | F1 score |
|---|---|---|---|---|
| *k*-means-$L^2$-PDF | **0.67** | 0.66 | **0.69** | **0.67** |
| *k*-means-$L^2$-CDF | 0.63 | 0.62 | 0.67 | 0.65 |

| Method | | | | |
|---|---|---|---|---|
| $k$-means-Euclidean | 0.62 | 0.62 | 0.60 | 0.61 |
| $k$-means-Manhattan | 0.65 | 0.66 | 0.60 | 0.63 |
| $k$-NN-$L^2$-PDF | 0.65 | **0.77** | 0.43 | 0.55 |
| $k$-NN-$L^2$-CDF | 0.63 | 0.65 | 0.57 | 0.61 |
| $k$-NN-Euclidean | 0.63 | 0.66 | 0.55 | 0.60 |
| $k$-NN-Manhattan | 0.61 | 0.69 | 0.40 | 0.50 |
| NSC | **0.67** | **0.72** | 0.56 | 0.63 |
| RF | 0.62 | 0.61 | 0.63 | 0.62 |
| GB | 0.60 | 0.60 | 0.63 | 0.61 |
| LASSO | 0.62 | 0.62 | 0.64 | 0.63 |
| RR | 0.64 | 0.63 | **0.67** | **0.65** |